\documentclass[twocolumn]{aastex631}

\shortauthors{Li et al.}
\graphicspath{{./}{figures/}}

\usepackage{amsmath}

\usepackage{CJKutf8}
\newcommand{\TLchinesename}{{\begin{CJK}{UTF8}{gbsn}(李坦达)\end{CJK}}}
\newcommand{\SLchinesename}{{\begin{CJK}{UTF8}{gbsn}(毕少兰)\end{CJK}}}
\newcommand{\ZLchinesename}{{\begin{CJK}{UTF8}{gbsn}(李志凯)\end{CJK}}}
\newcommand{\YZchinesename}{{\begin{CJK}{UTF8}{gbsn}(周一啸)\end{CJK}}}

\makeatletter
\newcommand*\fsize{\f@size pt\relax}
\makeatother
\usepackage{comment}

\newcommand{\Kepler}[0]{\emph{\textit{Kepler}}}

\newcommand{\logg}[0]{$\log g$}
\newcommand{\Dnu}[0]{$\Delta\nu$}
\newcommand{\dnu}[1]{$\delta\nu_{#1}$}
\newcommand{\numax}[0]{\mbox{$\nu_{\mathrm max}$}} 
\newcommand{\stagger}{\texttt{Stagger}}
\newcommand{\garstec}{\texttt{GARSTEC}}
\newcommand{\mean}[1]{\ensuremath{\langle #1 \rangle}}

\begin{document}
\title{Validating the 1D–3D coupling stellar models via Asteroseismology of 18 \Kepler{} main-sequence stars}
\newcommand{\BNUFrontiers}{Institute for Frontiers in Astronomy and Astrophysics, Beijing Normal University, Beijing, 102206, China}
\newcommand{\BNUSchool}{School of Physics and Astronomy, Beijing Normal University No.19, Xinjiekouwai St., Haidian District, Beijing, 100875, China}
\newcommand{\UC}{School of Physical and Chemical Sciences --- Te Kura Mat{\=u}, University of Canterbury, Private Bag 4800, Christchurch 8140, Aotearoa, New Zealand}
\newcommand{\UIO}{Rosseland Centre for Solar Physics, Institute of Theoretical Astrophysics, University of Oslo, P.O. Box 1029, Blindern, NO-0315 Oslo, Norway}

\author[0009-0008-9478-7807]{Li Zhikai\ZLchinesename}
\affiliation{\BNUFrontiers}
\affiliation{\BNUSchool}

\author[0000-0001-6396-2563]{Li Tanda\TLchinesename}
\affiliation{\BNUFrontiers}
\affiliation{\BNUSchool}
\correspondingauthor{Li Tanda}
\email{E-mail: litanda@bnu.edu.cn (LT)} 

\author[0000-0003-0817-6126]{Zhou Yixiao\YZchinesename}
\affiliation{\UC}
\affiliation{\UIO}

\author[0000-0002-7642-7583]{Bi Shaolan\SLchinesename}
\affiliation{\BNUFrontiers}
\affiliation{\BNUSchool}

\begin{abstract}
Standard 1D stellar evolution model has poor descriptions of the near-surface layers of stars, and this can be improved by using the atmosphere model computed from 3D hydrodynamical simulations. In this work, we validated the model inferences of the 1D–3D coupling models using 18 well-studied stars from the \textit{Kepler} LEGACY Sample. We compared our estimates of the fundamental parameters determined with other six pipelines and obtained good consistency. The results indicate that the 1D–3D coupling models can be applied to characterizing solar-like stars with confidence. Our analysis showed similar pattern for the surface term in stars with effective temperature range from $\sim$5000\,K to $\sim$6400\,K, suggesting that the surface term of the 1D–3D coupling models is simpler and easier to deal with than that of models using classical atmosphere.
\end{abstract}


\section{Introduction} \label{sec:introduction}
Stellar structure and evolution models have long been widely used to determine the fundamental parameters of stars. The method involves fitting theoretical models to observations to characterize stars \citep{2014ApJS..210....1C,2015MNRAS.452.2127S}. A long-standing issue in the stellar model is the poor description of the near-surface layers due to the simple treatment of convection and atmosphere \citep{2017LRCA....3....1K,refId0}. A direct consequence is that the oscillation mode frequencies calculated from the theoretical model are systematically offset from the actual observed values and the deviation increases with frequency. This discrepancy was first identified in the
Sun and later seen in all solar-like oscillators \citep{1988Natur.336..634C,1996Sci...272.1286C,1988A&A...200..213D,1996A&A...307..609A,2018MNRAS.478.4697B}, indicating that it is widespread among stars with convective envelopes. The systematic offsets between model and observed oscillation frequencies have been referred to as the surface effect \citep{1984Sci...226..687B} and several parameterization methods have been developed to correct it \citep{2008ApJ...683L.175K,2014A&A...568A.123B,2015A&A...583A.112S}. Although surface effect corrections bring theoretical frequencies into better agreement with measured values, it does not account for the impact of a simplified treatment of convection near the surface on stellar evolution \citep{2015A&A...577A..60S,2018MNRAS.478.5650M}.

A more robust solution that improves the simplified near-surface treatment in 1D stellar model is replacing it with realistic inputs from 3D hydrodynamic simulations. Early attempts to combine stellar interior models with 3D hydrodynamic simulations were mostly limited by the high computational cost and could only be applied to the Sun \citep{1997A&A...322..646S,1999A&A...351..689R}. In recent years, 3D hydrodynamic atmospheric simulation grids \citep{2009MmSAI..80..711L,2013ApJ...767...78T,2013ApJ...769...18T,2013A&A...557A..26M} have been developed to cover a wide range of stellar parameters and allows for the 3D atmospheric structure to replace the outer layers of 1D stellar structure model at different evolutionary stages. The model is referred to as a 1D–3D coupling model. 
Later on, the interpolation method proposed by \citet{2017MNRAS.472.3264J,2018MNRAS.481L..35J} enables 3D simulation data to be applied to compute the evolution of 1D stellar models across the HR diagram. The analysis of \citet{2021MNRAS.500.4277J} inferred that the 1D–3D coupling models predict global structural and evolutionary parameters of stars that deviate from the standard model. Specifically, under evolutionary tracks with the same mass, age, and metallicity, the difference in effective temperature between the two models can be as large as 80\,K, with stellar radius up to 25\% larger than that given by standard evolution calculations. \citet{2020MNRAS.491.1160M} applied the 1D–3D coupling models to asteroseismic studies and computed theoretical oscillation frequencies for several solar-like stars observed by the \textit{Kepler} mission\citep{2010Sci...327..977B}. Their results showed that the 1D–3D coupling models significantly improved the deviation between observed and theoretical oscillation frequencies.
However, validations of 1D–3D coupling models have been limited to a few targets at the solar metallicity, and their applicability to the determination of stellar fundamental parameters remains uncertain. Recently, \citet{10.1093/mnras/staf937} extended the interpolation scheme, enabling the application of 1D–3D coupling models beyond the solar metallicity regime. This advancement allows for testing the reliability in broader parameter ranges.

Well-studied asteroseismic targets are good benchmarks for validating the 1D–3D coupling models because mode frequencies of solar-like oscillations can characterize stars to high precision. \citet{2017ApJ...835..172L} selected 66 asteroseismic stars comprising the \textit{Kepler} LEGACY Sample and measured their individual oscillation frequencies. In a follow-up study, \citet{2017ApJ...835..173S} determined the stars' fundamental parameters using seven different modeling pipelines with varied input physics and numerical methods. Model inferences of these pipelines are ideal for testing the reliability of 1D–3D coupling model.

In this work, we aim to validate the 1D–3D coupling models by applying it onto the detailed modeling of the \textit{Kepler} LEGACY Sample and compare the inferred fundamental parameters to those determined with other pipelines.
In Section~\ref{Stellar Models and Grid}, we describe the construction of our 1D–3D coupling stellar models and the computational grid. Section~\ref{Target Selection and Model Fitting} details the sample selection and the method used to match models with observational constraints. The results of our analysis are presented in Section~\ref{result}, and in Section~\ref{conclusions}, we summarize and discuss the outcomes of this work.

\section{Stellar Models and Grid}
\label{Stellar Models and Grid}

Asteroseismic analysis is inevitably influenced by surface effect. However, the recently developed 1D--3D coupling stellar models have shown sufficient potential in compensating for the near-surface deficiencies of 1D stellar models. Most 1D models approximate the near-surface structure using the mixing-length theory (MLT). In addition to the errors inherent in this approximation, the choice of $\alpha_{\mathrm{MLT}}$ also introduces systematic uncertainties. The 1D–3D coupling approach replaces the near-surface convection layers in the stellar evolution simulation with a more realistic 3D hydrodynamic simulation atmosphere. Since the remaining deep convection regime is close to adiabatic, its structure is not sensitive to changes in the mixing-length parameter. Changing $\alpha_{\mathrm{MLT}}$ hardly alters the evolutionary tracks, and hence, we can fix the $\alpha_{\mathrm{MLT}}$ value in model computation \citep{10.1093/mnras/staf937}. This not only reduces the systematic uncertainties in modeling interference but also reduce the computation expence by eliminating one dimension of free input parameter when constructing model grids. Furthermore, compared with standard stellar evolution simulations that use Eddington or 1D model atmospheres to determine the outer boundary condition, the outer boundary condition used in the 1D–3D coupling approach is arguably more realistic.

\subsection{1D–3D Coupling Method}
\label{models}
In 1D stellar evolution codes, equations of stellar structure are closed by boundary conditions. The outer boundary is typically set at or near the photosphere, where physical quantities such as pressure are often obtained by integrating the Eddington gray atmosphere or from pre-computed model atmospheres. In the 1D–3D coupling approach, the near-surface region of the 1D models is replaced by the mean 3D (or \mean{\rm 3D}) models. Specifically, the stellar structural calculation only extends to the near-adiabatic convective layer below the stellar surface (0.14\%$R_{\odot}$ below the surface in the solar case), while the structure above is supplied from horizontal- and time-averaged, interpolated 3D models.

Models of stellar interior are computed using the Garching Stellar Evolution Code (\garstec{}, \citealt{2008Ap&SS.316...99W}). Input physics and numerical details of the evolution calculation are nearly identical to those used in \citet{10.1093/mnras/staf937}. 
We use the \texttt{FreeEOS} equation of state\footnote{Available at \url{http://freeeos.sourceforge.net/documentation.html}} \citep{2004Irwin...feos1,2012ascl.soft11002I} and OPAL opacity tables \citep{1996ApJ...464..943I}. Both equation of state and opacity data are calculated based on the \citet{2009ARA&A..47..481A} metal mixture.
Convection in \garstec{} is treated with the mixing-length theory in the formulation of \citet{kippenhahn2012stellar}, where the value of the mixing length parameter, $\alpha_{\rm MLT} = 2.77$, obtained from solar calibration with the 1D–3D coupling method, is fixed throughout the study. \citet{10.1093/mnras/staf937} demonstrated that for 1D–3D coupled models, a 20\% change of $\alpha_{\rm MLT}$ shifts the effective temperature by less than 20 K in main-sequence evolution.
Atomic diffusion for elements H, He, C, N, O, Ne, Mg, Si, and Fe is calculated following the method of \citet{1994ApJ...421..828T}. As atomic diffusion without radiative levitation results in strong sedimentation of helium and metals in warmer, F-type stars, which is not seen in observations, we include turbulent diffusion in the formulation prescribed by \citet[their Eq.~1]{2017ApJ...840...99D} to encounter the over-depletion of elements heavier than hydrogen.
Convective overshoot and mass loss are not considered in our models.

The outer boundary conditions of stellar interior models are provided by the \stagger{}-grid \citep{2013A&A...557A..26M,2024A&A...688A.212R}, a grid of more than 220 3D model atmospheres computed using the \stagger{}-code \citep{1995...Staggercodepaper,2018MNRAS.475.3369C,2024ApJ...970...24S}. The grid covers a wide range of stellar parameters, with $T_{\rm eff}$ ranging from $3500$ to $7000$ K, $\log (g/{\rm [cm/s^2]})$ from $1.5$ to $5$, and [Fe/H] from $-4$ to $0.5$, corresponding to F, G, K-type dwarfs and red giants from extremely metal-poor to metal-rich.

All models are constructed with detailed radiative transfer calculations but without magnetic fields. Vertically, the simulations span the lower atmosphere, the photosphere, and the super-adiabatic convective layers just below the surface. The bottom boundary of the simulation domain extends down to Rosseland mean opacity $\tau_{\rm Ross} \sim 10^6 - 10^7$, where the temperature gradient is close to adiabatic. The horizontal extent of simulations is determined based on the typical size of granules at corresponding stellar parameters.
3D radiative-hydrodynamical simulations are averaged in horizontal plane and time, yielding mean 3D structures ready for interpolation. We refer the readers to \citet{2013A&A...557A..26M} and \citet{2024A&A...688A.212R} for detailed descriptions of the construction of 3D models and \stagger{}-grid, and \citet{2017MNRAS.472.3264J} and \citet{10.1093/mnras/staf937} for details about the interpolation technique.

The essence of the 1D–3D coupling method is that the outer boundary conditions for stellar structure calculations are supplied by \mean{\rm 3D} models at every time step (see \citealt{2018MNRAS.481L..35J,2020MNRAS.491.1160M,10.1093/mnras/staf937} for more details about the method). The matching point (or the outer boundary of the stellar interior model) resides in the convective region below the surface, where the temperature gradient is close to adiabatic, while maintaining some distance from the bottom of the simulation domain to avoid numerical artifacts. Gravitational accelerations at the matching point are assumed to equal the surface gravity. 
At a given evolution time step, the temperature at the matching point from the 1D model, $T_{\rm m,1D}$, is used to derive the effective temperature by interpolating the function $T_{\rm eff}(T_{\rm m}, \log g, {\rm [Fe/H]})$ defined by mean 3D models. The thus obtained surface parameters are used in interpolation for the corresponding $\mean{\rm 3D}$ structure. Outer boundary condition of the stellar structural calculation requires the pressure of the $\mean{\rm 3D}$ and 1D model to be equal at the matching point. The second boundary condition is that luminosity evaluated from the Stefan–Boltzmann law, $L = 4 \pi R^2 \sigma T_{\rm eff}^4$, equals the luminosity given by \garstec{}, where $R$ is the photospheric radius determined by integrating the \mean{\rm 3D} model on top of the 1D model and $\sigma$ is the Stefan-Boltzmann constant. The numerical solver in \garstec{} will iteratively adjust the interior structure until the two conditions are fulfilled.

\subsection{Computation of Models}
\label{Computation of Models}

We constructed a grid of models using the aforementioned 1D–3D coupling models. The model grid covers a mass range of 0.8--1.4\,$M_\odot$ and a metallicity range of [Fe/H] = -0.3 to 0.3. Initial masses and metallicities of the 1000 tracks are distributed in the parameter space following the \citet{SOBOL196786} sampling. The initial helium mass fraction is determined via a helium enrichment law $Y = 1.05Z + 0.248$, where the gradient is adopted from \citet{2021MNRAS.505.2427L}.
We start our computation at the Hayashi line and terminate the evolution at the end of main sequence (when the centre Hydrogen fraction goes below 0.0001). Theoretical oscillation frequencies were then computed based on the structure models using the \textsc{GYRE} Stellar Oscillation Code \citep{2018MNRAS.475..879T}.
These frequencies are used in Section~\ref{CNM} to match against asteroseismic data, serving as constraints for model selection.

\section{Target Selection and Model Fitting}
\label{Target Selection and Model Fitting}
\subsection{Target Selection}
\label{sample}
\begin{figure}
    \includegraphics[width=\columnwidth]{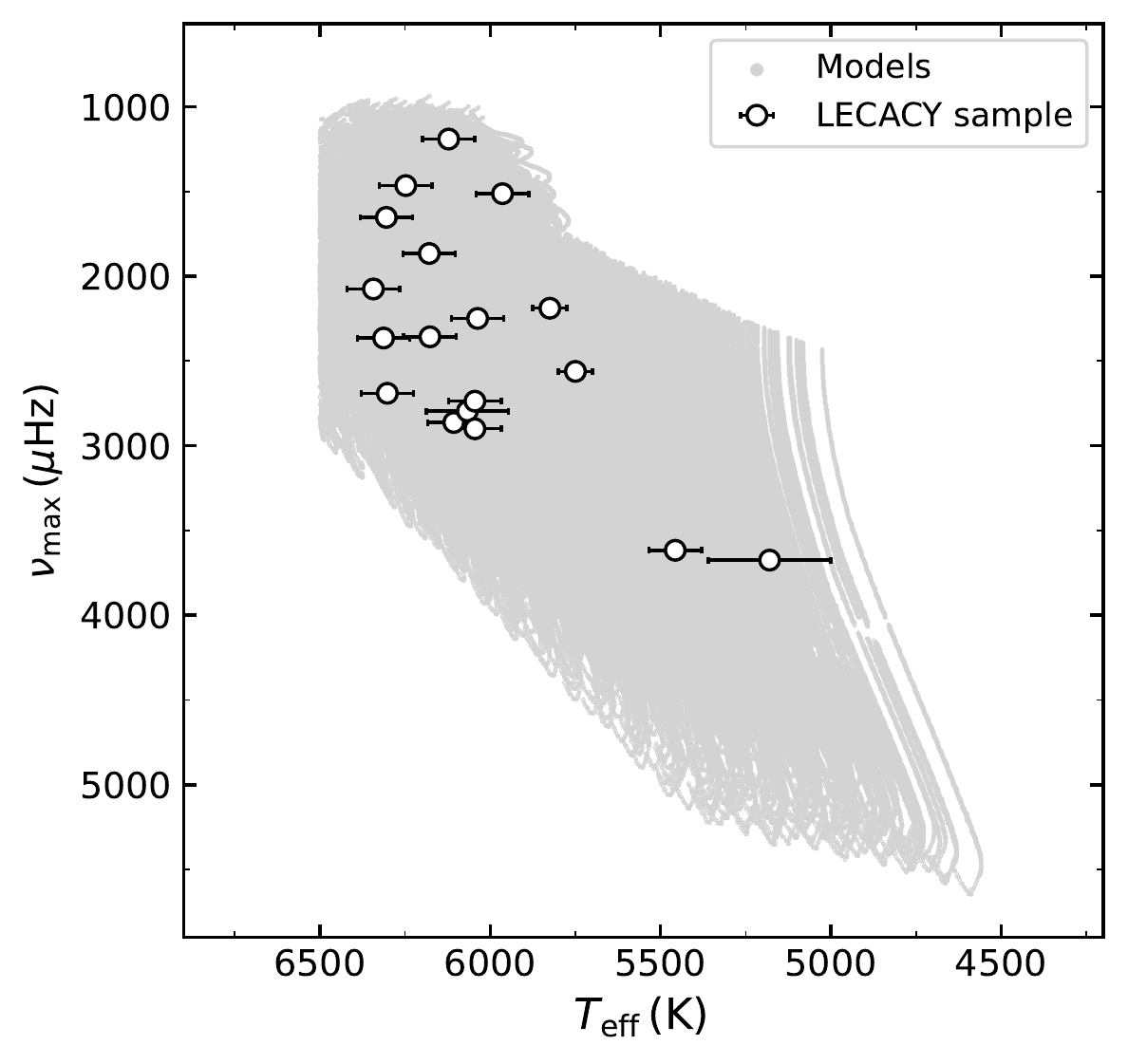}
    \caption{The computed stellar models and selected star sample on the $T_{\mathrm{eff}}$-$\nu_{\mathrm max}$ diagram. Gray dots represent stellar evolutionary tracks computed with the 1D–3D coupling model. 
    Open circles indicate the 18 selected stars.}
    \label{HRd}
\end{figure}

We select our star sample from the 66 main-sequence stars that comprise the \textit{Kepler} LEGACY sample, as studied by \citet{2017ApJ...835..173S}. We select stars with $T_{\mathrm{eff}}$ below 6500\,K, since the grid of 3D atmospheric models becomes sparse above this temperature, making interpolated results less reliable. Given the typical metallicity uncertainty of $\sigma_{\mathrm{[Fe/H]}}\approx 0.10\,\mathrm{dex}$, we restrict the sample to $-0.1<\mathrm{[Fe/H]}<+0.1$, thereby retaining a $\geq 3\sigma$ sampling margin within our model grid. Lastly, stars locate near the edges of model parameter ranges ($T_{\mathrm{eff}}$ and $\nu_{\mathrm{obs}}$) are excluded to avoid the edge effects. 
We finally obtain 18 main-sequence stars as demonstrated in Figure~\ref{HRd}.

\subsection{Fitting Scheme}
\label{CNM}

We applied the fitting method proposed by \citet{2023MNRAS.523...80L} in our analysis. The method is based on the Gaussian Process (GP) kernels and describes the systematic errors and uncertainties in theoretical model frequencies as correlated noise models (CNM). The main reason to adopt this fitting scheme is that the GP kernels are flexible to treat the surface term. Previous studies have shown that the 1D–3D coupling models can present different patterns of the surface term from standard models \citep{2018MNRAS.481L..35J,2020MNRAS.491.1160M}. Moreover, the function forms have not been widely tested in many stars. In this case, the methods developed for standard models may not be suitable to correct for this surface term. 

In brief, \citet{2023MNRAS.523...80L} described the systematic offset in oscillation frequencies by introducing a mean function into a modified GP kernel, and considered the systematic uncertainty introduced by the resolution of the model grid with two GP kernels.
These prior offsets are incorporated, in the form of three GP kernels, into the likelihood of the observed and theoretical model oscillation frequencies. The likelihood function is optional and we use the multivariate Student’s $t$-distribution because it takes into account potential mismeasured mode frequencies. This allows us to evaluate the match between observed and theoretical model oscillation frequencies while accounting for systematic offset, and to select the best-fitting model. The resulting likelihood values are then used as weights in the corner plot to statistically infer the stellar mass, age, and radius (see Figure~\ref{coner}). 

\begin{figure*}
    \centering
    \includegraphics[width=\textwidth]{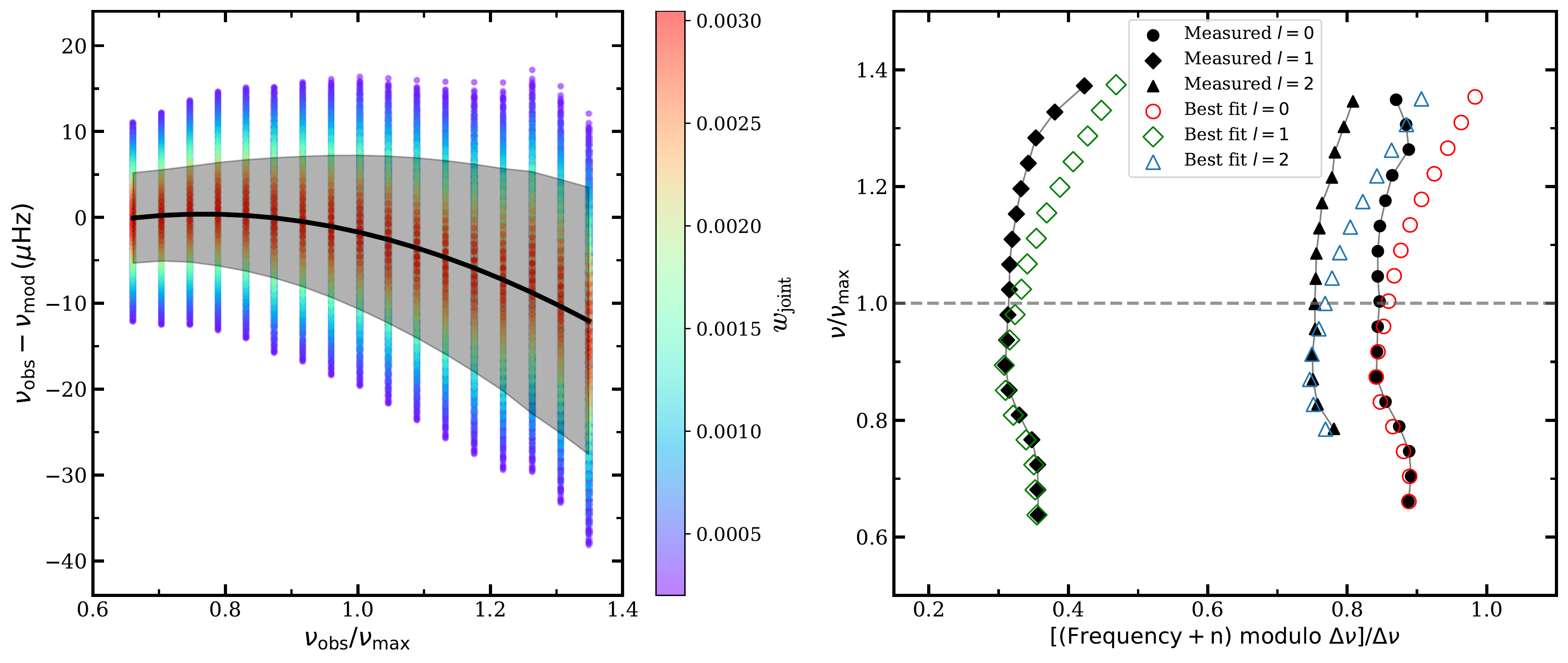}
    \caption{Left: Determination of the mean function ($\mu_{\varepsilon}$)  and variation ($\sigma_{\varepsilon}$)  of the systematic error kernel ($\varepsilon$) for KIC 8379927. Dots are frequency differences between observed and theoretical models. The colour code indicates the joint weight derived from multiple likelihood functions. The solid line represents the polynomial function that fits the frequency differences (weighted by the joint weight), and we use this as the mean function. The grey shade indicates the weighted standard deviation which is adopted as the variance. Right: \'Echelle diagram for KIC 8379927 illustrating the comparison between observed and the best-fitting model across three different radial orders. The x-axis shows frequencies shifted by a constant \textit{n} to ensure radial ridge continuity. Black dots represent the observed individual frequencies, while colored open circles denote the corresponding theoretical frequencies from the best-fitting model. Gray solid lines connect observed modes of the same angular degree for clarity, and the gray dashed line indicates the position of $\nu_{\mathrm{max}}$.}
    \label{lk}
\end{figure*}

Throughout the entire fitting procedure, we only modified the method used to determine the mean function. \citet{2023MNRAS.523...80L} introduced a joint weight $\left(w_{\mathrm{joint}}=w_{\mathrm{low}\text{-}\nu} \cdot w_{\nu_{\mathrm{max}}} \right)$ based on the priors of frequency deviation below 0.7\numax{} and that at \numax{} to find the mean function. For the 1D–3D coupling models, the amount and the function form of frequency deviations are mostly unknown and the method is hence not suitable. \citet{2003A&A...411..215R} tested how changes in the outer layers of 1D stellar models affect the ratio of small to large separations of acoustic oscillations and showed that this ratio remains insensitive to the outer layers and effectively characterizes the internal structure. Therefore, we introduce the frequency ratio $\delta\nu_{02}/\Delta\nu$ and other global parameters into the joint weight $w_{\mathrm{joint}}$ to determine the mean function, written as 
\[
   w_{\mathrm{joint}} = w_{\mathrm{low}-\nu} \cdot w_{T_{\mathrm{eff}}} \cdot w_{\mathrm{[Fe/H]}} \cdot w_{\Delta\nu} \cdot w_{\delta\nu_{02}/\Delta\nu}
\]
To directly compare with the model inferences of these stars, we use the same observed constraints including effective temperature, metallically, seismic global parameters (\Dnu{}, and \dnu{02}), and oscillation frequencies provided by \citet{2017ApJ...835..173S} and \citet{2017ApJ...835..172L}.
We demonstrate our fitting results of an example star in Figure~\ref{lk}.
As shown in the plot on the left, the black solid line represents the resulting mean function $\mu_{\varepsilon}$. The weighted standard deviation of the residuals is used as the noise variance $\sigma_{\varepsilon}$ in the systematic error kernel, shown as the gray-shaded region. The best-fitting model is illustrated on the so-called \'Echelle diagram in the right panel.

\section{result}
\label{result}

\subsection{Modeling Inferences}

\begin{figure}
    \vspace{2\baselineskip}  
    \includegraphics[width=0.47\textwidth]{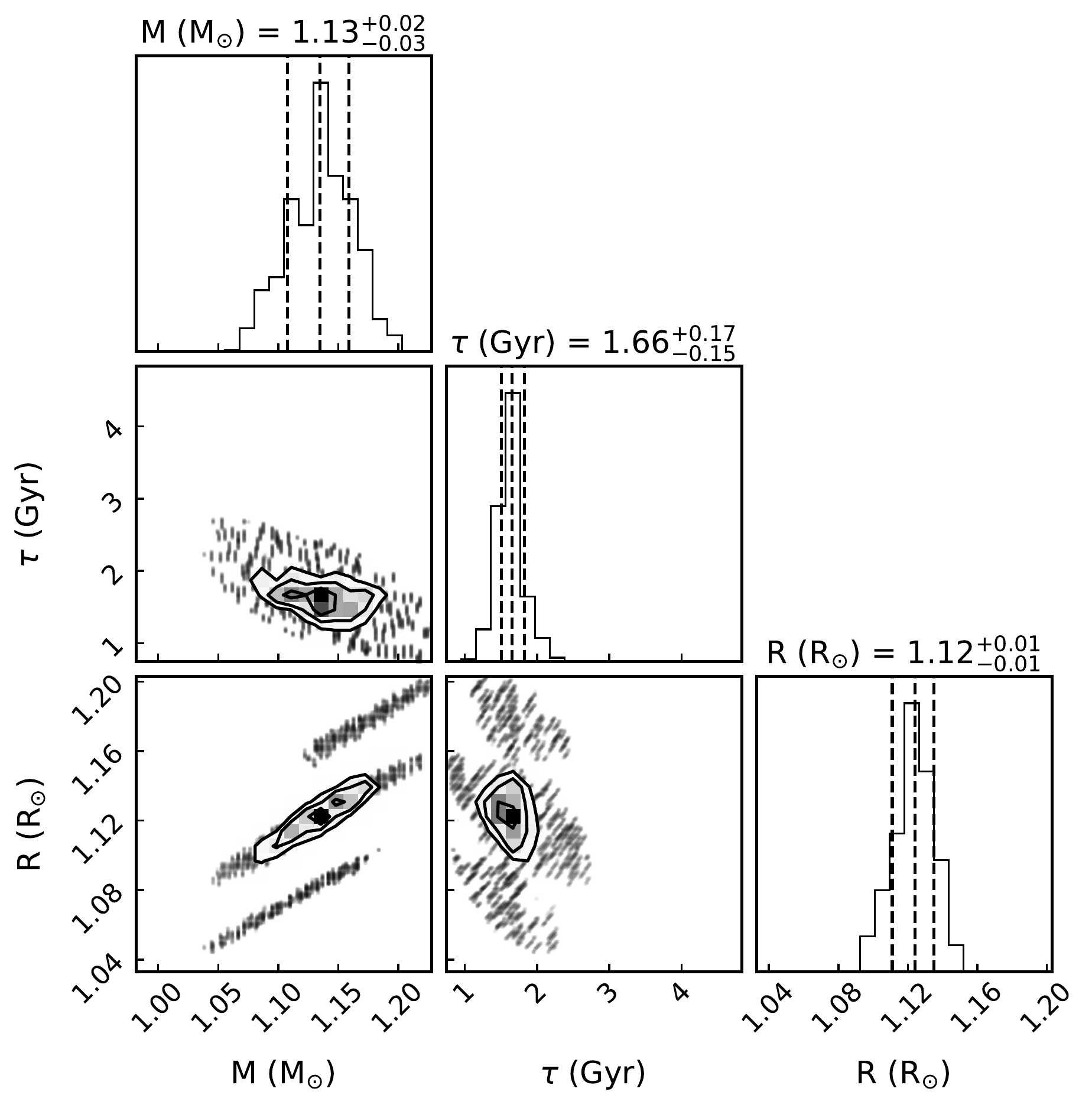}
    \caption{Probability distributions of mass, age, and radius for KIC 8379927 on the \textsc{corner} plot \citep{Foreman-Mackey2016}.}
    \label{coner}
\end{figure}

\renewcommand{\arraystretch}{1.1}
\begin{table}
\centering
\caption{Fundamental parameters for the 18 \textit{Kepler} LEGACY stars determined with the 1D–3D coupling model.}
\begin{tabular}{lccc}
\hline
KIC & Mass ($M_\odot$) & Age (Gyr) & Radius ($R_\odot$) \\
\hline
3427720  & $1.097^{+0.028}_{-0.027}$ & $2.408^{+0.124}_{-0.165}$ & $1.112^{+0.012}_{-0.012}$ \\
3735871  & $1.132^{+0.032}_{-0.031}$ & $1.685^{+0.301}_{-0.294}$ & $1.108^{+0.012}_{-0.013}$ \\
6106415  & $1.111^{+0.016}_{-0.020}$ & $4.701^{+0.225}_{-0.094}$ & $1.232^{+0.008}_{-0.009}$ \\
6225718  & $1.156^{+0.029}_{-0.025}$ & $2.409^{+0.147}_{-0.215}$ & $1.229^{+0.014}_{-0.012}$ \\
7206837  & $1.380^{+0.004}_{-0.010}$ & $1.907^{+0.100}_{-0.087}$ & $1.571^{+0.003}_{-0.006}$ \\
7771282  & $1.203^{+0.013}_{-0.009}$ & $4.365^{+0.279}_{-0.191}$ & $1.611^{+0.011}_{-0.009}$ \\
8179536  & $1.271^{+0.026}_{-0.038}$ & $1.948^{+0.273}_{-0.294}$ & $1.360^{+0.012}_{-0.018}$ \\
8228742  & $1.347^{+0.024}_{-0.011}$ & $3.618^{+0.102}_{-0.190}$ & $1.855^{+0.011}_{-0.003}$ \\
8379927  & $1.134^{+0.024}_{-0.027}$ & $1.655^{+0.170}_{-0.148}$ & $1.124^{+0.011}_{-0.013}$ \\
9139151  & $1.119^{+0.024}_{-0.028}$ & $2.030^{+0.443}_{-0.163}$ & $1.135^{+0.011}_{-0.011}$ \\
9955598  & $0.919^{+0.011}_{-0.019}$ & $6.507^{+0.336}_{-0.310}$ & $0.893^{+0.004}_{-0.006}$ \\
10454113 & $1.245^{+0.017}_{-0.041}$ & $1.614^{+0.238}_{-0.226}$ & $1.270^{+0.006}_{-0.019}$ \\
10644253 & $1.146^{+0.027}_{-0.027}$ & $1.341^{+0.178}_{-0.153}$ & $1.112^{+0.011}_{-0.012}$ \\
11772920 & $0.860^{+0.016}_{-0.019}$ & $9.040^{+0.582}_{-0.569}$ & $0.857^{+0.006}_{-0.007}$ \\
12009504 & $1.190^{+0.030}_{-0.021}$ & $4.002^{+0.232}_{-0.360}$ & $1.406^{+0.013}_{-0.010}$ \\
12069424 & $1.059^{+0.016}_{-0.010}$ & $7.120^{+0.464}_{-0.222}$ & $1.220^{+0.006}_{-0.005}$ \\
12069449 & $1.015^{+0.014}_{-0.012}$ & $6.936^{+0.337}_{-0.171}$ & $1.106^{+0.005}_{-0.004}$ \\
12258514 & $1.221^{+0.005}_{-0.014}$ & $5.197^{+0.033}_{-0.146}$ & $1.589^{+0.004}_{-0.007}$ \\
\hline
\end{tabular}
\label{tab:stellar_parameters}
\end{table}

\begin{figure*}
    \centering
    \includegraphics[width=\textwidth]{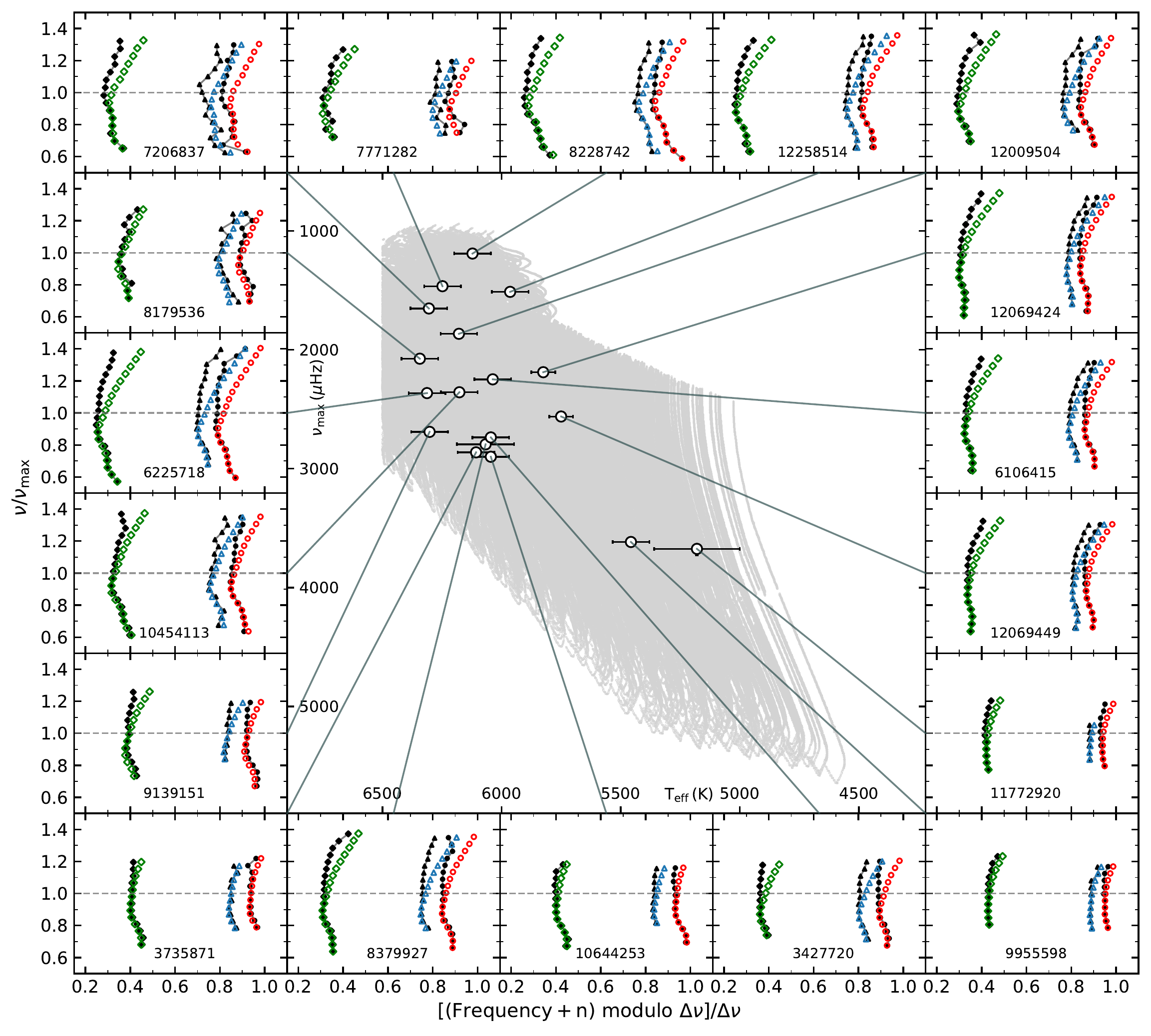}
    \caption{Center: Asteroseismic Hertzsprung-Russell diagram, identical to Figure~\ref{HRd}, showing the positions of the 18 \textit{Kepler} LEGACY stars. Surroundings: \'Echelle diagrams for the 18 \textit{Kepler} LEGACY stars, illustrating the comparison between observed oscillation modes and the best-fitting models across three different radial orders. The symbols and lines follow the same convention as in the right panel of Figure~\ref{lk}}
    \label{HRd2}
\end{figure*}

\begin{figure*}
    \centering
    \includegraphics[width=\textwidth]{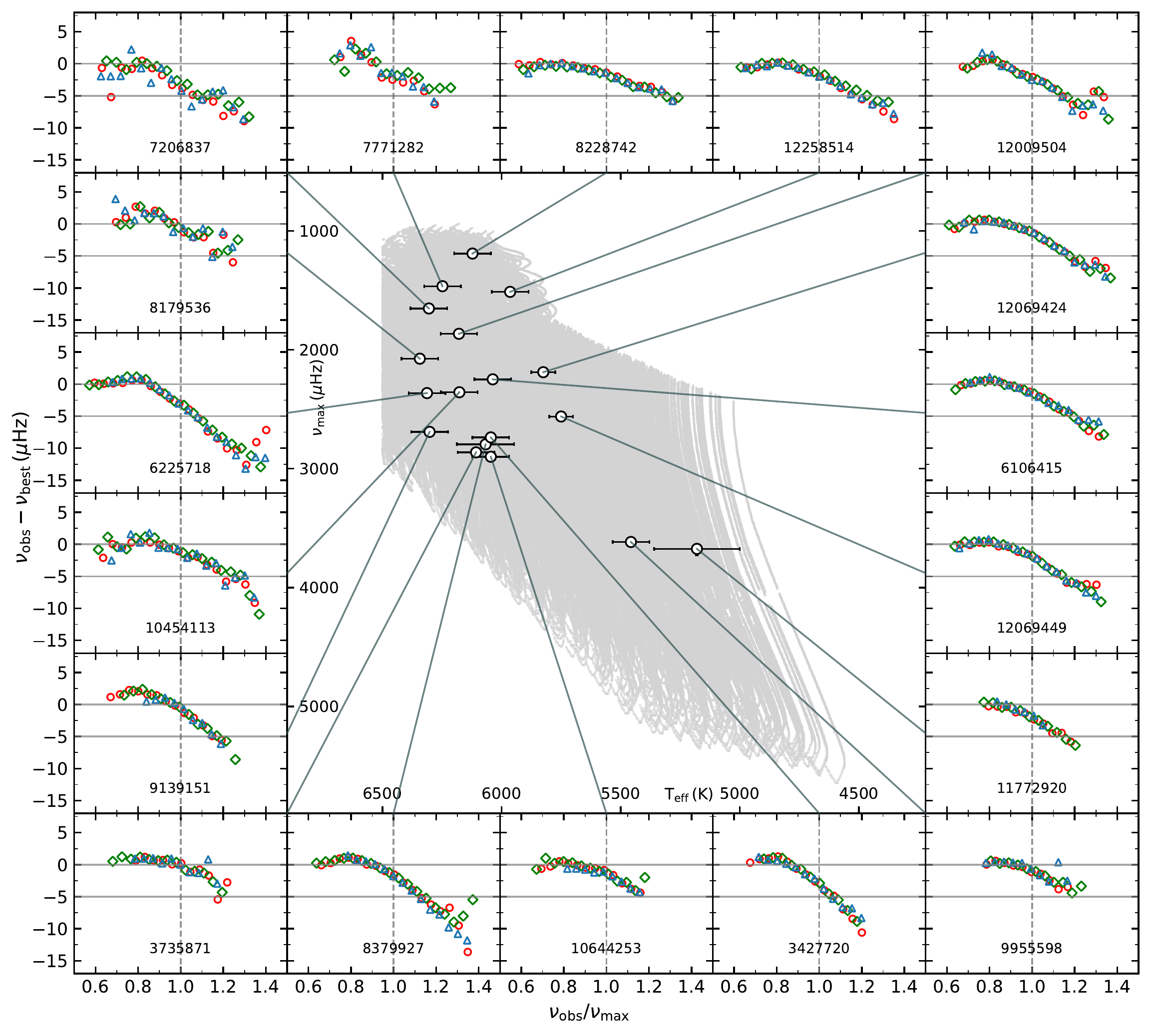}
    \caption{Center: Asteroseismic HR diagram, identical to Figure~\ref{HRd}, showing the positions of the 18 \textit{Kepler} LEGACY stars. Surroundings: Frequency differences between the best-fitting model frequencies and the observed oscillation frequencies for each star. The x-axis for each panel is centered at the star’s $\nu_{\mathrm{max}}$ (indicated by the gray dashed line). The gray horizontal solid lines mark the positions of 0 and $5\mu$Hz on the y-axis. A darkslategray solid line connects the position of each star in the asteroseismic HR diagram to its corresponding frequency-difference panel.}
    \label{HRd1}
\end{figure*}

We applied the aforementioned fitting scheme in modeling the 18 selected main-sequence stars. We adopted the \textsc{corner} tool \citep{Foreman-Mackey2016} to examine the likelihood distribution for each star. We use the median of probability distribution as the estimated value, and the 16th and 84th percentile to determine the upper and lower uncertainties for a certain parameter. 
Figure~\ref{coner} illustrates the likelihood distributions and model inferences of mass, age, and radius for an example star. 
Results for all 18 stars are summarized in Table~\ref{tab:stellar_parameters}.

Figure~\ref{HRd2} displays the positions of 18 stars from the \textit{Kepler} LEGACY sample in the Asteroseismic HR diagram at the center, while the surrounding panels show comparisons with the best-fitting models in the form of \'Echelle diagrams. When fitting the mean function, we included the weighting factor $w_{\mathrm{low}\text{-}\nu}$ to enhance the influence of models that better match the observations at low frequencies. In the \'Echelle diagrams, we can observe that the magnitude of frequency deviations due to surface effect increases with frequency. This trend was confirmed during the determination of the mean function representing the systematic surface frequency offset. Additionally, our theoretical asteroseismic frequencies fail to reproduce certain observed helium glitch signature in some targets, and this is because the helium fraction is not an independent parameter in our model inputs.

\begin{figure*}
    \includegraphics[width=\textwidth]{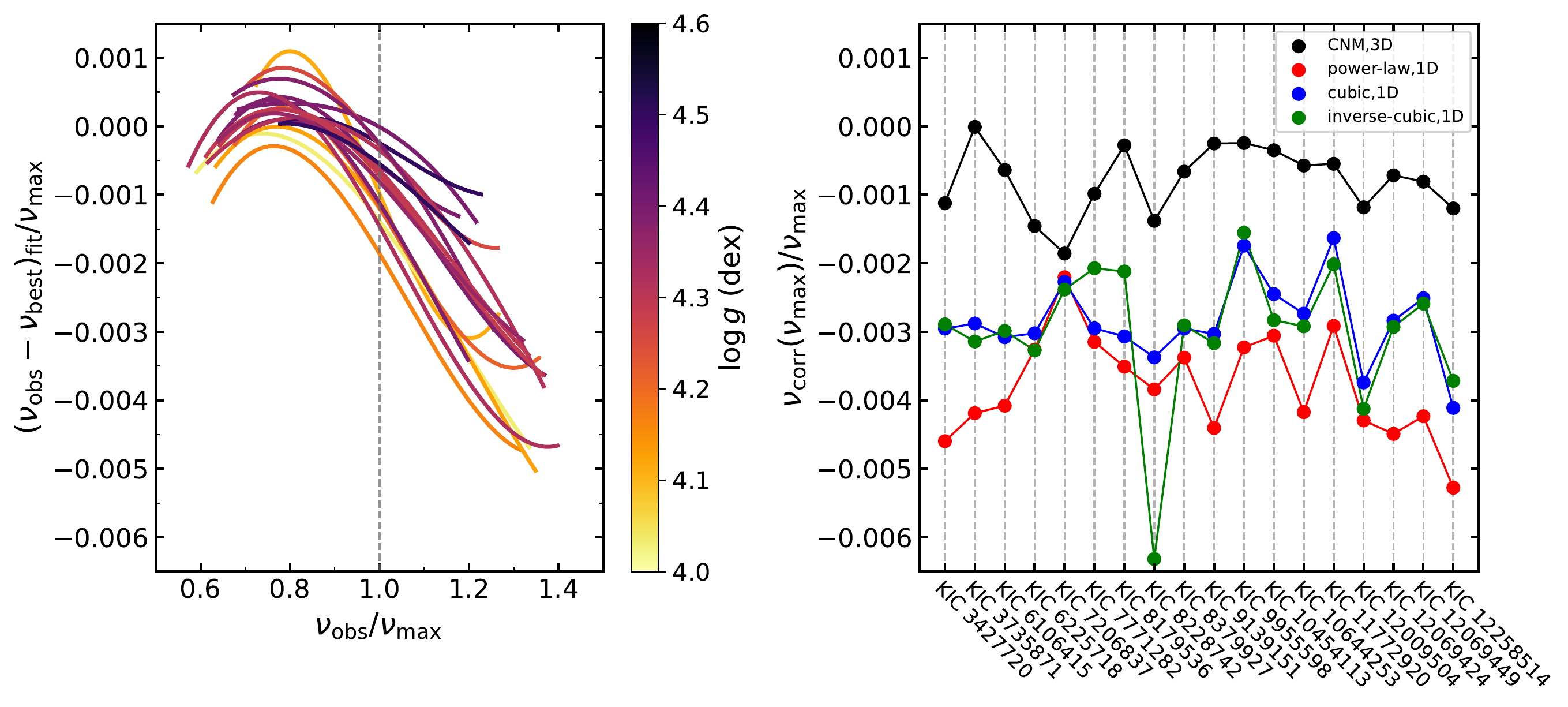} 
    \caption{Left: The cubic function fit results for the frequency differences of the radial modes of 18 stars. Both the x-axis and y-axis are normalized to $\nu_{\mathrm{max}}$, and the color bar represents the $\log g$ values for each source. The gray dashed line indicates the position of $\nu_{\mathrm{max}}$. Right: Comparison of relative surface corrections at $\nu_{\mathrm{max}}$ for the 18 \textit{Kepler} LEGACY stars. The black points represent the surface corrections obtained using the CNM method in our 1D-3D coupling model, while the colored points represent the surface corrections obtained using the power-law correction (red), the cubic method (blue), and the inverse-cubic term method (green) for the 1D models by \citet{2018MNRAS.479.4416C}.}

    \label{nfd}    
\end{figure*}

\subsection{The Surface Terms in 18 Stars}

The surface term of the 1D–3D coupling models have been examined in the Sun and a few solar-like stars. The frequency deviations do not monotonically depends on the observed frequency like those of the standard 1D model. The offsets between the observation and the model ($\nu_{\mathrm{obs}}$ - $\nu_{\mathrm{mod}}$) slightly increase with observed frequency from almost zero to a small positive value below 0.7\numax{}, and decrease with observed frequency above 0.7\numax{}. The typical deviation found in previous works ranges from a few $\mu$Hz to approximately $6\,\mu$Hz \citep{2018MNRAS.481L..35J,2020MNRAS.491.1160M}. 

We demonstrate the surface term (systematic frequency offset between observations and the best-fitting model) of the 18 stars in Figure~\ref{HRd1}. The frequency offsets in the surrounding panels clearly indicate that the surface term also exhibits a structure with a slight rise at low frequencies and a gradual decline at higher frequencies. In other words,  the surface term is not a simple power law across the entire oscillation frequency range.

Moreover, for each star in Figure~\ref{HRd1}, the maximum offset between the observed frequencies and the theoretical frequencies of the 1D–3D coupling models is below \( 15 \, \mu\mathrm{Hz} \), resulting in smaller surface term compared to the 1D models. Due to the presence of convective turbulence, 3D simulations shift the photosphere and oscillation cavity slightly upwards compared to 1D models, as illustrated in Figure 4 of \citet{2018MNRAS.481L..35J} and was also discussed in previous works such as \citet{2016A&A...592A..24M}. The slight expansion of oscillation cavity reduces the p-mode frequencies, bringing the theroetical frequencies from the 1D-3D coupling models closer to the measured values. The effect is more pronounced in modes with higher frequencies as they have more substantial amplitude near the surface. 

We normalize the frequency offsets of the 18 stars to \numax{}, as shown in the left panel of Figure~\ref{nfd}. The surface terms of most stars exhibit an anti-correlation with \logg \, and follow a similar functional form. This is different from standard models, whose surface terms are temperature dependent, presenting a smooth function with frequencies in G-type stars and a relatively wiggled function form in F-type stars \citep[see Figure~8,][]{2018MNRAS.479.4416C}. 
Additionally, in the right panel of Figure~\ref{nfd}, we compare the frequency offsets at \numax{} obtained from the CNM, as described in Section~\ref{CNM}, in the 1D–3D coupling models with those from \citet{2018MNRAS.479.4416C}, who used three surface correction methods in the 1D models. the frequency offsets from the 1D–3D coupling models at \numax{} are within the range of $-0.002$\numax{}. These deviations are generally smaller than those found in 1D models ($-0.01$ to $-0.002$\numax{} corresponding to $\log g$ from 4.0 to 4.5 dex).

As for the remaining systematic offset, part of them arise from the breakdown of the adiabatic approximation in the near-surface layers when computing stellar oscillations. Additionally, our models only include modifications to the density stratification by patching to \mean{\rm 3D} models, whereas contributions from the turbulent pressure and the coupling between turbulence and oscillation are not taken into account \citep{2009LRSP....6....2N,2015LRSP...12....8H}. Considering these factors, we conclude that the 1D–3D coupling models, constructed by interpolating 3D atmosphere grids into 1D models, effectively reduce the systematic offset introduced by 1D convection approximations, with the interpolation error remaining within a reasonable range. However, surface corrections are still necessary when fitting individual oscillation frequencies.

\subsection{Validating the 1D–3D Coupling Model}
\begin{figure*}
    \includegraphics[width=\textwidth]{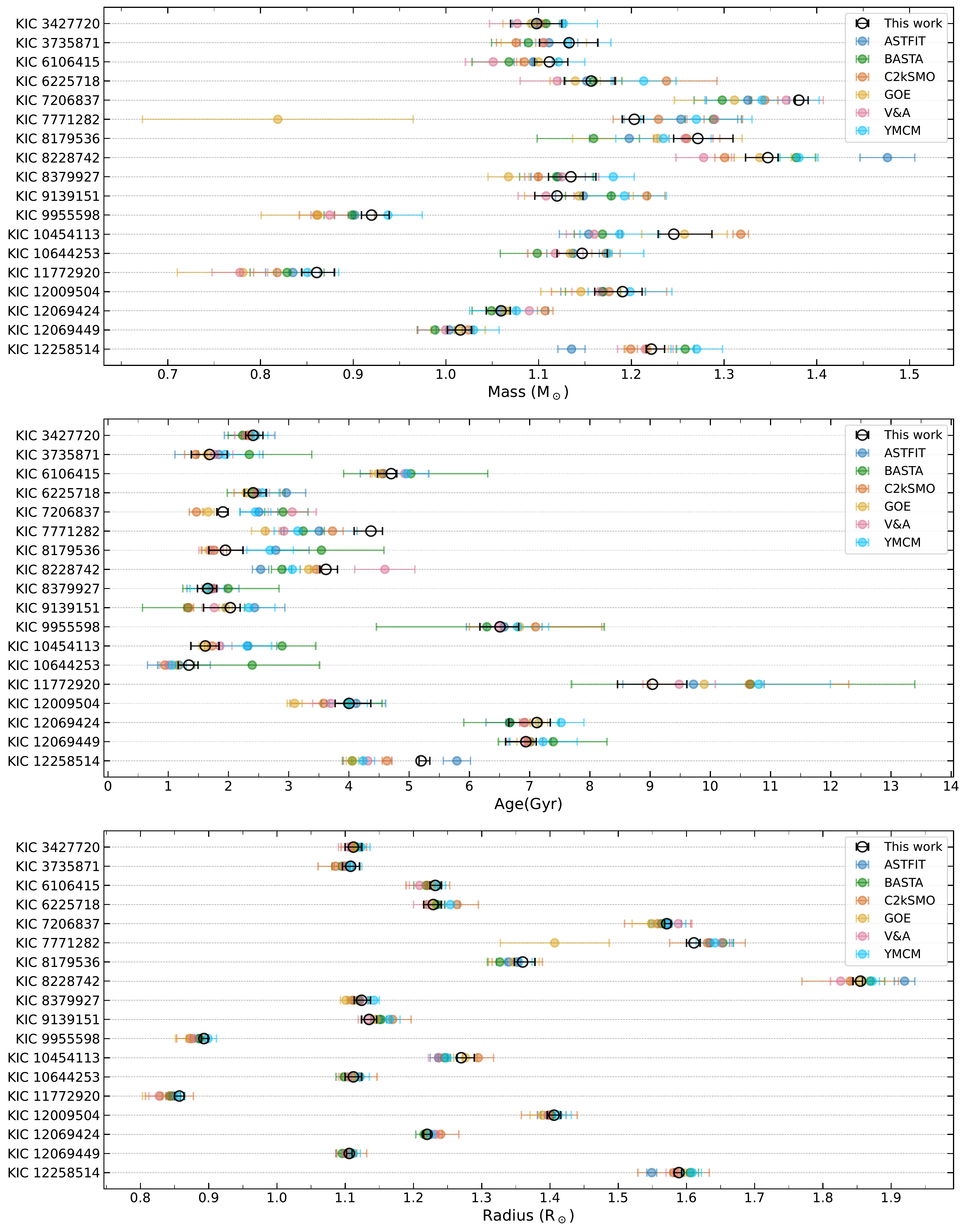} 
    \caption{The mass, age, and radius results for the 18 \textit{Kepler} LEGACY stars. The black open circles with error bars represent the stellar parameters determined using the 1D–3D coupling models. The colored circles with error bars show the stellar parameters derived by multiple asteroseismic modeling pipelines, as reported by \citet{2017ApJ...835..173S}.}
    \label{properties}    
\end{figure*}

\begin{figure*}
    \includegraphics[width=\textwidth]{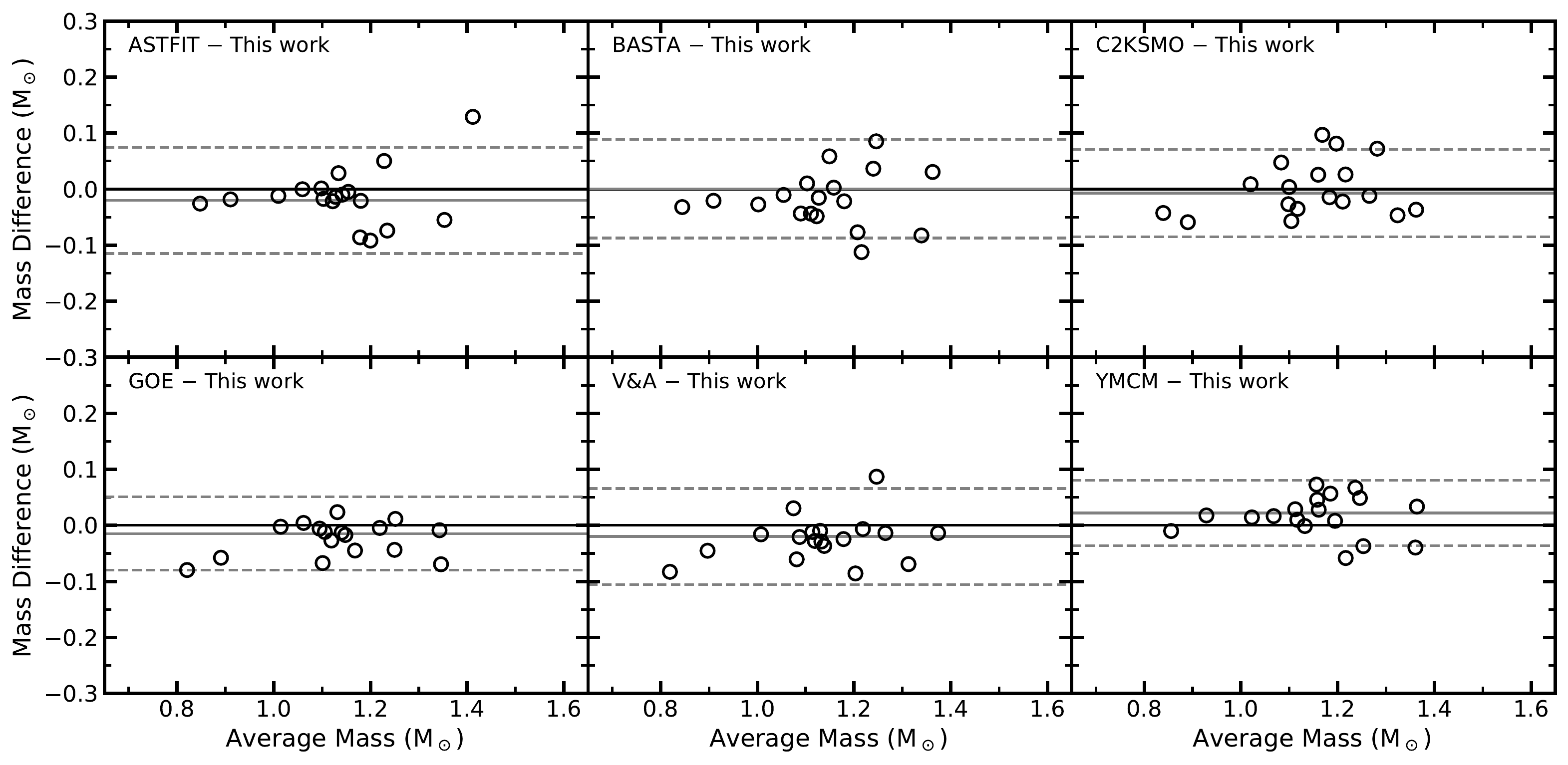} 
    \caption{Comparison of mass results from different pipelines with this work. The x-axis shows the average mass between each pipeline and this work, and the y-axis displays the difference between the pipeline and this work. Solid black lines indicate the zero level for reference. The solid and dashed grey lines represent the weighted mean difference and twice the standard deviation, respectively, with uncertainties combined in quadrature. }
    \label{mass}    
\end{figure*}
\begin{figure*}
    \includegraphics[width=\textwidth]{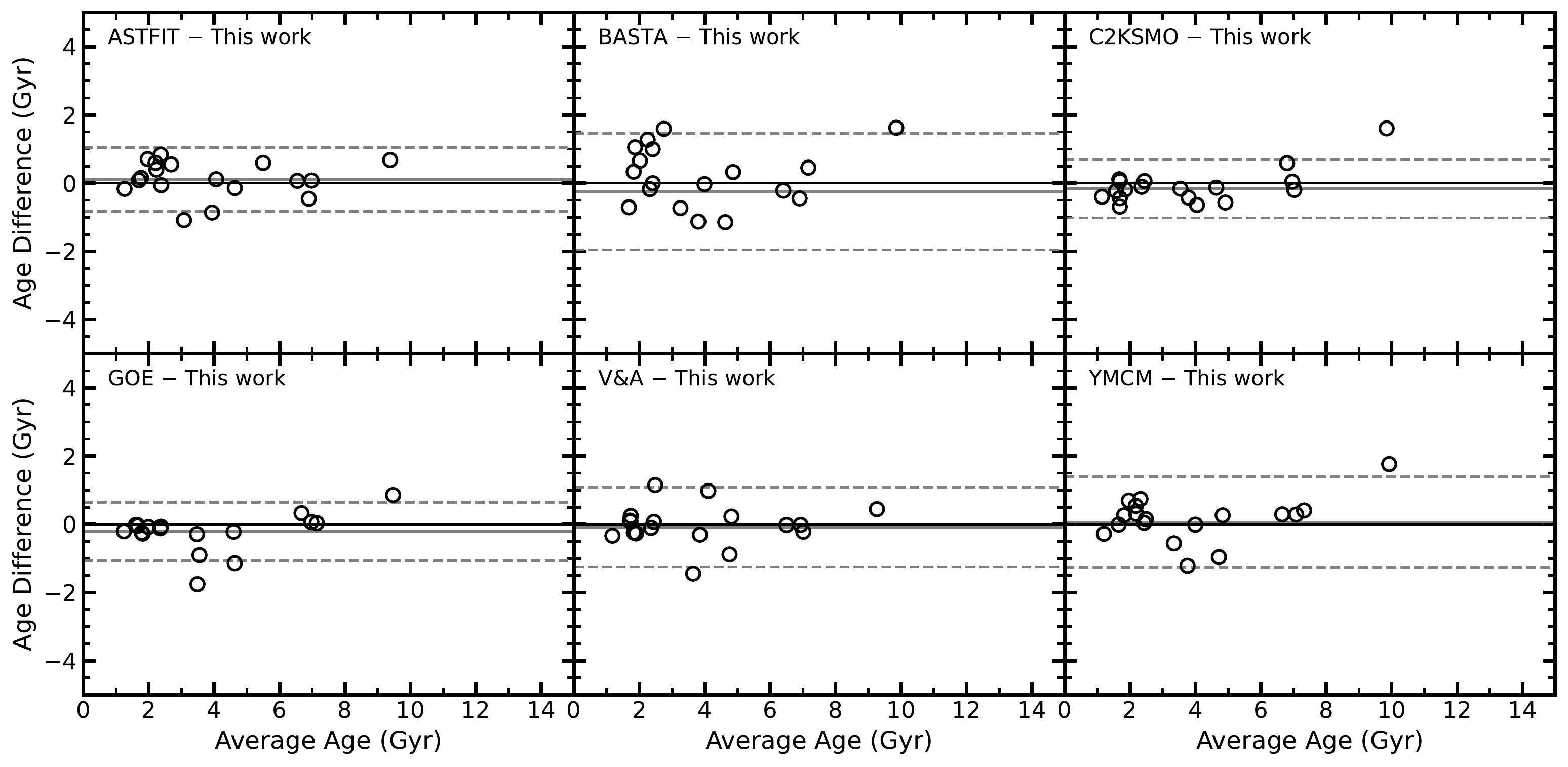} 
    \caption{Age results from different pipelines compared to this work. All other features are the same as in Figure~\ref{mass}}
    \label{age}    
\end{figure*}
\begin{figure*}[!htbp]
    \centering
    \includegraphics[width=\textwidth]{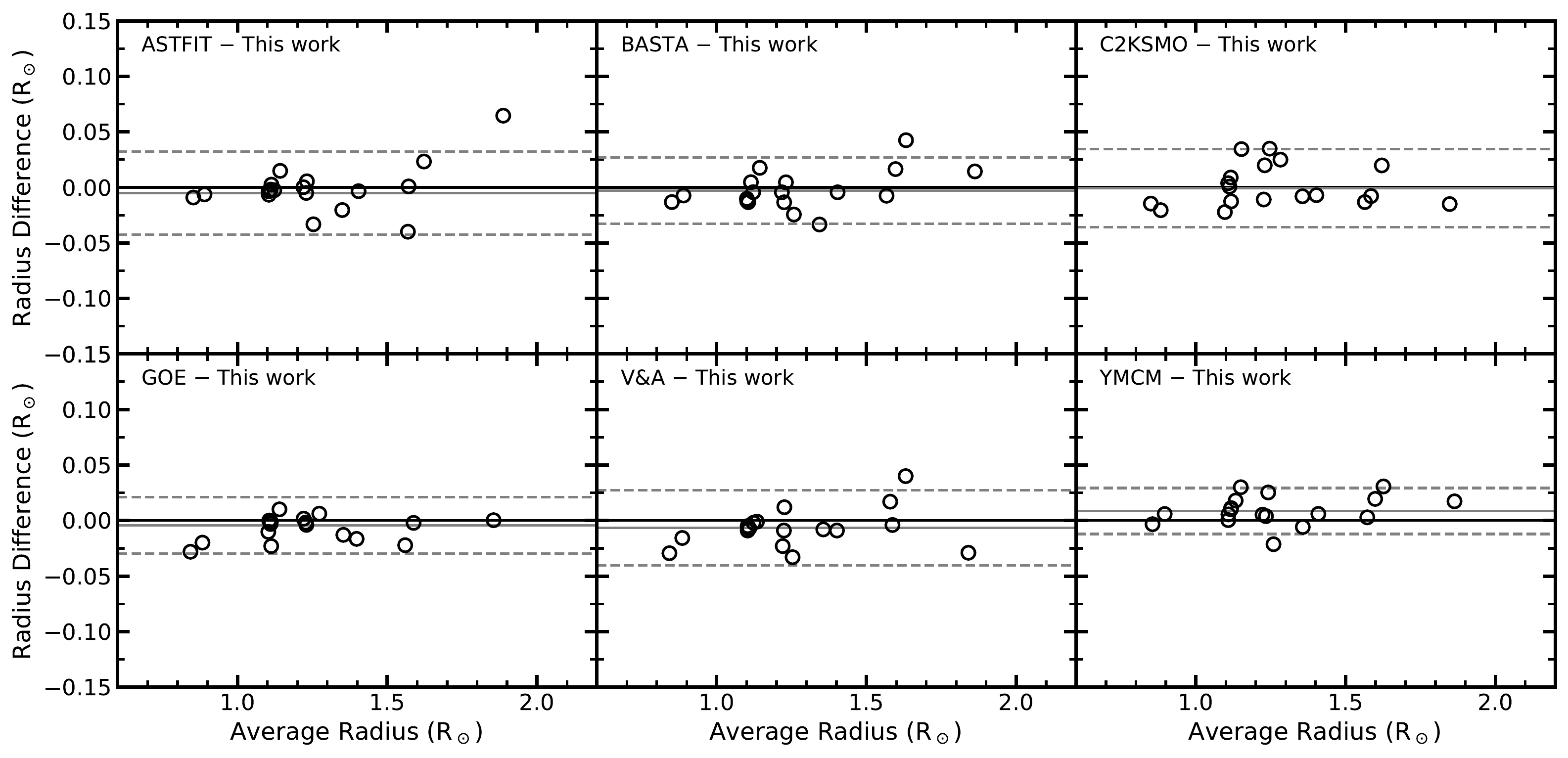} 
    \caption{Radius results from different pipelines compared to this work. All other features are the same as in Figure~\ref{mass}}
    \label{radius}    
\end{figure*}

Accurately determining stellar fundamental parameters is key to assessing the reliability of stellar models. One way to validate model inferences is to compare them with independent measurements, such as dynamical masses from binaries or stellar radius from interferometry. In addition, parallaxes inferred from asteroseismic distances \citep[via asteroseismic radius and angular diameters,][]{2012ApJ...757...99S} can be compared with trigonometric parallaxes to assess the accuracy of the asteroseismic radius scale. Although Gaia delivers high-precision parallaxes, most techniques for measuring stellar fundamental parameters remain observationally demanding and are typically limited to relatively bright, nearby targets. \citet{2017ApJ...835..173S} compared several independent determinations with those from asteroseismic modeling pipelines based on 1D models for the \textit{Kepler} LEGACY sample, showing very good agreement. We therefore do not repeat such comparisons for the 1D–3D coupling models. Instead, we treat the results from different the pipelines as independent measurements, select six of them, and adopt them as representative stellar parameters. We then compare these with the results from our 1D–3D coupling models to assess their accuracy and reliability.

We carry out two types of comparison of the stellar fundamental parameters: first, a comparison for each star across all models, as shown in Figure~\ref{properties}, to assess the precision of the 1D–3D coupling models for individual targets; and second, a comparison for each model across all stars, as shown in Figures~\ref{mass}-\ref{radius}, to examine the statistical agreement between the 1D–3D coupling models and the 1D models.

Figure~\ref{properties} presents the stellar fundamental parameters—mass, age, and radius—for the 18 stars in the \textit{Kepler} LEGACY sample, as determined using the 1D–3D coupling models. In each panel, the vertical axis lists the 18 targets, while the horizontal axis shows the corresponding values of the stellar parameters. For comparison, we also show the results from six independent teams that employed asteroseismic modeling pipelines based on 1D stellar models. As illustrated in Figure~\ref{properties}, the black open symbols with error bars (representing the 1D–3D coupling model results) are entirely enclosed within the ranges of the colored circles with error bars corresponding to the 1D model results. This demonstrates that the 1D–3D coupling models yield results that are consistent with those of 1D models within the stated uncertainties, indicating that they can achieve equally accurate determinations of stellar fundamental parameters.

In addition, we perform a statistical comparison between the results of the 1D models and the 1D–3D coupling models, in order to assess the statistical consistency of the stellar fundamental parameters. The comparison is conducted by computing the differences relative to each of the 1D model results. Figure~\ref{mass}, Figure~\ref{age}, and Figure~\ref{radius} show the comparisons for mass, age, and radius, respectively. In the figures, we indicate the weighted mean and twice the standard deviation of the differences, with weights based on the uncertainties of the stellar parameters. The mass differences are within 0.1 $M_{\odot}$, the age differences are within 2 Gyr, and the radius differences are within 0.05 $R_{\odot}$. The weighted mean, shown as the gray solid line, is essentially consistent with zero, indicating that the stellar parameters derived from the 1D–3D coupling models are statistically consistent with those from the 1D models.

Both types of comparison demonstrate that the stellar fundamental parameters obtained from the 1D–3D coupling models are consistent with those from the 1D models. This holds true whether the focus is on individual stars or on the statistical properties of the entire sample. These results indicate that the stellar parameters derived from the 1D–3D coupling models are comparably accurate to those inferred from the 1D models.

\section{Conclusions}
\label{conclusions}

We have presented a validation of the 1D–3D coupling stellar models through asteroseismic analysis of 18 main-sequence stars from the \textit{Kepler} LEGACY sample. Using a grid constructed from the 1D–3D coupling method, which provides a more physically realistic description of the near-surface convective region, we determined the stellar fundamental parameters of these stars. A comparison between the standard 1D models and the 1D–3D coupling models shows a good level of agreement in the derived stellar parameters. The 1D–3D coupling models are thus capable of providing reliable stellar parameters, enabling their broader application to solar-like stars.

By comparing the best-fitting 1D–3D coupling stellar models with the observations, we find that the structural surface effect are significantly reduced relative to the 1D models, and the remaining terms share a common functional form, demonstrating the strong potential of the 1D–3D coupling models for asteroseismic applications.

In summary, compared to standard 1D models, the 1D–3D coupling stellar models provide a more realistic description of the near-surface region, reducing uncertainties caused by the choice of the mixing length parameter and effectively eliminating a dimension in grid-based modeling.
Model grids constructed from the 1D–3D coupling approach now cover a wide range of stellar parameters and, like other stellar models, are sufficient to represent real stellar properties.
This enables robust comparisons with observations through asteroseismology and allows accurate stellar parameters to be derived. In the era of upcoming space missions, such as PLATO and ET, the 1D–3D coupling method
can thus serve as a powerful stellar modeling tool,
supporting more extensive studies in stellar physics.

\section*{Acknowledgments}

This work is supported by the National Natural Science Foundation of China (NSFC) (grants 12373031, 12090040/12090042, 12588202), the National Key R\&D Program of China No.2023YFE0107800, No.2024YFA1611900, No. 2024YFA1611901, China Manned Spaced Program with grant no.CMS-CSST-2025-A12, and the Fundamental Research Funds for the Central Universities.
YZ gratefully acknowledges support from the Elaine P. Snowden Fellowship. 
YZ acknowledges support from the European Union's Horizon 2020 research and innovation programme under the Marie Skłodowska–Curie grant agreement No 101150921.
The numerical results presented in this work were partly obtained at the Centre for Scientific Computing, Aarhus \url{https:// phys.au.dk/forskning/faciliteter/cscaa/}.

This paper includes data collected by the Kepler mission and obtained from the MAST data archive at the Space Telescope Science Institute (STScI). Funding for the Kepler mission is provided by the NASA Science Mission Directorate. STScI is operated by the Association of Universities for Research in Astronomy, Inc., under NASA contract NAS 5–26555.

\bibliographystyle{aasjournal}  
\bibliography{Article}  
\end{document}